# Interactive query expansion for professional search applications

Tony Russell-Rose, Goldsmiths, University of London, 25 St James's, London, England, SE14 6AD T.Russell-Rose@gold.ac.uk

Philip Gooch, Scholarcy, 71-75 Shelton Street Covent Garden, London, England, WC2H 9JQ phil@scholarcy.com

Udo Kruschwitz, Universität Regensburg, Universitätsstraße 31, D-93053 Regensburg, Germany Udo.Kruschwitz@ur.de

## Abstract

Knowledge workers (such as healthcare information professionals, patent agents and recruitment professionals) undertake work tasks where search forms a core part of their duties. In these instances, the search task is often complex and time-consuming and requires specialist expert knowledge to formulate accurate search strategies. Interactive features such as query expansion can play a key role in supporting these tasks. However, generating query suggestions within a professional search context requires that consideration be given to the specialist, structured nature of the search strategies they employ. In this paper, we investigate a variety of query expansion methods applied to a collection of Boolean search strategies used in a variety of real-world professional search tasks. The results demonstrate the utility of context-free distributional language models and the value of using linguistic cues such as ngram order to optimise the balance between precision and recall.

**Keywords**: professional search, query expansion, information retrieval, natural language processing, machine learning, ontologies

# 1. Introduction

Many knowledge workers rely on the effective use of search applications in the course of their professional duties (Verberne et al., 2019). For example, healthcare information professionals perform systematic searching of published literature sources as the foundation of evidence-based medicine (Russell-Rose & Chamberlain, 2017). Likewise, patent agents rely on prior art search as the foundation of their due diligence process (Lupu et al., 2011). Similarly, recruitment professionals use Boolean search as the foundation of the candidate sourcing process (Russell-Rose & Chamberlain, 2016a).

However, systematic literature reviews can take years to complete (Bastian et al., 2010), and new research findings may be published in the interim, leading to a lack of currency and potential for inaccuracy (Shojania et al., 2007). Likewise, patent infringement suits have been filed at a rate of more than 10 a day due to the later discovery of prior art which their original search missed (Gibbs, 2006). And recruitment professionals report that finding candidates with appropriate skills and experience continues to be their primary concern (Russell-Rose & Chamberlain, 2016b). Each of these domains has its own expert competencies and communities of practice, but conceptually they share a need to execute searches that are comprehensive, transparent and reproducible (Mullins et al., 2014). It is this common need that motivates the work described in this paper.

There is another motivation for our work, and that is the discrepancy between academic research in information systems and actual industry use cases. It has been pointed out that evaluation in academic projects tends to focus on idealised tasks that are less complex than those found in industry and that investigating *realistic* use cases is a fundamental step in bridging the gap between academia and industry (Karlgren, 2019). We see our work as a contribution toward this goal.

**Fig 1**. The World Health Organisation's Clinical Trials Search Portal

The traditional solution to structured search problems is to use form-based query builders such as that shown in Figure 1. The output of these tools is typically a series of Boolean expressions consisting of keywords, operators and ontology terms, which are combined to form a multi-line artefact known as a *search strategy* (Figure 2).

```
1   A01N0025-004/CPC
2   RODENT OR RAT OR RATS OR MOUSE OR MICE
3   BAIT OR POISON
4   2 AND 3
5   1 OR 4
6   AVERSIVE OR ADVERSIVE OR DETER? OR REPEL?
7   NONTARGET OR (NON WITH TARGET) OR HUMAN OR DOMESTIC OR PET
OR DOG OR CAT
8   6 AND 7
9   8 AND 5
10  BITREX OR DENATONIUM OR BITREXENE OR BITTERANT OR BITTER
11  10 AND 5
12  9 OR 11
```

**Fig. 2** An example patent search strategy

In this paper, we review the role of query expansion within the context of professional structured search strategies. We investigate a number of techniques for generating interactive query suggestions, and evaluate them using a variety of real-world data. The guiding principle in our evaluation is to provide replicable experiments that will also serve as a benchmark for future investigations.

## 2. Background

### 2.1 Professional search

The term 'professional search' refers to search for information in a work context which often involves complex information needs, the use of multiple repositories and the incorporation of domain-specific taxonomies or vocabularies (Verberne et al., 2018) or a combination of different relevance criteria (Jiaming Qu et al., 2020). Various authors have provided descriptive and behavioral definitions of the term (see (Russell-Rose et al., 2018) for an overview). One of the earliest definitions was proposed by Koster et al. (Koster et al., 2009), whereby professional search:

- Is performed by a professional for financial compensation;
- Is within a particular domain and/or area of expertise;
- Has a specified brief, which is typically well defined but complex;
- Has a high value outcome where the results will reduce risk, provide assurances, etc.;
- Has budgetary constraints such as time and money.

A key distinction between professional search tasks and other kinds of search tasks, such as casual search (Elsweiler et al., 2012) and web search[1] (Broder, 2002) is that the latter:

- Are typically performed on a discretionary basis;
- Are not necessarily performed by an expert searcher or domain expert;
- And do not place at stake the professional reputation of the searcher.

There is a long history of study into how professionals search in Boolean environments (e.g. (Hersh et al., 2001)). However, professional search gained renewed momentum around a decade ago with the introduction of the TREC Legal Track which focused on e-discovery (Baron et al., 2006), followed later by the TREC Total Recall Track (Grossman et al., 2016). In recent years there has been a renewed interest in systematic literature searching, both from a theoretical (Scells et al., 2020) and practical perspective (Scells & Zuccon, 2018).

Given the complexity of professional search tasks and their reliance on specialist terminology, query expansion offers a natural approach to assist the searcher (Liu et al., 2011). Query expansion is the process of reformulating or augmenting a user's query in order to increase its effectiveness (Manning et al., 2008). Many web search engines, for example, offer query expansion in the form of auto-complete suggestions. Ruthven found, however, that searchers can have difficulty in identifying useful terms for effective expansion (Ruthven, 2003). Despite this, query suggestions can still be useful, as they can help in the search process even if they are not actively selected (Kelly et al., 2009).

---

[1] However, some professional search could be mediated via the web, and conversely, not all work-based searching is professional in nature

## 2.2 Query Expansion

The primary methods for query expansion are referred to as either *local* (based on documents retrieved by the query) or *global* (using resources independent of the query). Selection of suggested expansion terms can be either *automated* (applied without explicit user interaction) or *interactive* (guided by the user).

Global methods involve the use of resources such as thesauri, controlled vocabularies or ontologies to identify related terms in the form of synonyms, hypernyms, hyponyms, etc. (Aggarwal & Buitelaar, 2012). Such resources may be either *manually curated* or generated from text corpora using *distributional methods*. Automated global methods can increase recall significantly but may also reduce precision by adding irrelevant or out-of-domain terms to the query (Manning et al., 2008).

Ontologies are more useful for query expansion when they are specific to the task domain. Generic resources such as WordNet are considered less useful and may not distinguish class concepts from instances (Bhogal et al., 2007). Some ontologies offer an additional source of related terms in the form of words occurring in the term definitions (Navigli & Velardi, 2003). In the biomedical domain, expanding queries with related MeSH terms has been shown to be useful (Rivas et al., 2014), while adding synonyms from the more comprehensive UMLS has been found to improve recall (Griffon et al., 2012), at the expense of precision (Zeng et al., 2012). Query expansion in this context can actually benefit from incorporating a range of domain-specific knowledge bases rather than simply tapping into a single source, e.g. (Balaneshinkordan & Kotov, 2019).

The development of efficient distributional methods has revolutionized unsupervised natural language processing techniques for finding related terms (Collobert et al., 2011; T Mikolov et al., 2013). Consequently, a number of researchers have considered the utility of word embeddings for query expansion. Kuzi (Kuzi et al., 2016), Roy (Roy et al., 2016) and Diaz (Diaz et al., 2016) all used local embeddings trained on TREC corpora, with differing results. While Kuzi (Kuzi et al., 2016) found that local word embeddings outperformed the standard RM3 relevance model, Roy (Roy et al., 2016) found the opposite. Diaz (Diaz et al., 2016) compared local embeddings (TREC corpus) with global (generic Gigaword corpus) and found that local embeddings provided significantly better results for query expansion than global embeddings. More recently, we have seen that contextual embeddings, such as those based on BERT, have transformed the state of the art not only in natural language processing (Devlin et al., 2019) but also in information retrieval (Lin, 2019; Mitra & Craswell, 2018). However, given the nature of our research where we expand query terms on an individual basis, we will focus on context-free embeddings in our experiments.

A fundamental problem with most query expansion techniques is that queries may be harmed as well as improved (Xiong & Callan, 2015). In addition, with fully automated techniques the user may be unable to control how the expansion terms are applied. Moreover, Cao et al (Cao et al., 2008) argue that previous work considers only the effect of a complete set of expansion terms on retrieval, and ignores the issue of how to distinguish useful expansion terms from non-useful terms, e.g. as explored in (Gooda Sahib et al., 2010). We address these issues by treating query expansion as a *recommendation* task, i.e. given a query term entered by the user, can we recommend further relevant terms. Framing the task in this way is significant, since the use of an interactive approach allows the user to exercise a more informed judgement regarding both term selection and application within a structured search strategy. More broadly, this approach aligns with the goal of offering state-of-the-art query support in professional search while preserving transparency and interpretability (J Qu et al., 2021).

This approach also reflects a broader evolution among search systems from simple lookup tasks to more complex, exploratory, information seeking behaviours (White & Roth, 2009; White, 2016).

## 2.3 Application Context

Query suggestions are a common feature of many web search engines, and have served as the focus of many research studies e.g. (Efthimiadis, 1996; Tahery & Farzi, 2020). Since search queries on the web typically consist of short sequences of keywords with little or no linguistic structure (Beitzel et al., 2007; Kumar et al., 2020), term suggestions can offer immediate value as either an addition to the current query or as a wholesale replacement (Kruschwitz et al., 2013).

Although there have been studies investigating query expansion within a professional search context, e.g. Kim et al (Kim et al., 2011), Verberne et al (Verberne et al., 2014), Verberne et al (Verberne et al., 2016), examples of commercial systems in production are relatively rare. This may be due in part to the challenges presented by the structured nature of the queries themselves. For example, when sourcing candidates for a client brief, recruiters might use a structured query such as that shown in Figure 3.

```
Java AND (Design OR develop OR code OR
Program) AND ("* Engineer" OR MTS OR "*
Develop*" OR Scientist OR technologist) AND
(J2EE OR Struts OR Spring) AND (Algorithm OR
"Data Structure" OR PS OR "Problem Solving")
```

**Fig 3**. An example recruitment search strategy

For a query such as this, it is no longer sufficient to offer suggested terms as simple additions or as wholesale replacements. Instead, term suggestions must not only be relevant, but also specific to the structured nature of the query and the individual subexpressions it contains. In the above example, query suggestions relevant to the first subexpression would be quite inappropriate for the second subexpression.

In addition, the professional search context introduces a number of other important considerations regarding the evaluation process:

- Many use cases are oriented towards high-recall, set retrieval tasks (Tait, 2014), so evaluation methods based on the relevance ranking of search results are less appropriate.
- The suggested terms are scoped to a subexpression within a larger search strategy, so the evaluation must consider the specific context of each subexpression.
- Professional searchers may wish to select and apply expansion terms individually, so the evaluation should consider the contribution of each term individually rather than the effect of a candidate set as a whole.

We have therefore structured our evaluation using an approach based on previous query suggestion studies (Albakour et al., 2011), (Adeyanju et al., 2012), in which existing, human-generated resources are treated as a 'gold standard'. In this context, the task of the query suggestion system is to predict steps in a sequence, e.g. queries submitted by a user in the context of past interactions. Gold standard resources for this are often sampled from query logs (where available). This is in principle similar to evaluating chatbot responses, e.g. (Y. Wu et al., 2019), or news recommendation systems, e.g. (F. Wu et al., 2020), using logged interactions for evaluation purposes. In our case, a gold standard exists in the form of published search strategies. In this context, the evaluation process measures the extent to which terms found in those strategies can be predicted[2]. For example, given the term *rodent* in line 2 of the strategy of Figure 2, we measure the extent to which the related terms *rat*, *rats*, *mouse*, and *mice* can be predicted. This particular example contains five such disjunctions (lines 2, 3, 6, 7 and 10), so it offers five opportunities for evaluation. Moreover, since we use publicly available sources (rather than, for example, proprietary log data) our experiments can be more easily replicated by others.[3]

---

[2] Note that this approach constitutes a very strict evaluation procedure since terms labelled as false positives may in fact be true positives in a live task scenario (which we review in the Discussion).

Arguably, an ideal test collection for such an evaluation would contain search strategies curated specifically for the purpose. However, whilst such a resource may prove necessary, it may not be sufficient. For example, an ideal test collection should also include:

- Search strategies which are actively maintained and updated by the professional community (as opposed to purely archival collections)
- Search strategies from more than one domain, to allow investigation of the extent to which domain-specific resources will generalise to other domains.

There is no single collection that meets all three criteria of being curated, in current use, and cross-domain. For our test collection we therefore aggregated samples from the following resources:

1. The CLEF 2017 eHealth Lab (Goeuriot et al., 2017) is an evaluation initiative which includes a curated set of 20 topics for Diagnostic Test Accuracy (DTA) reviews. Each of these topics includes a manually constructed search strategy created by subject matter experts. The 20 search strategies in this collection yielded 102 disjunctions containing 898 terms (i.e. a mean of 8.80 terms per disjunction). Each term consists of a mean of 1.40 tokens.
2. The SIGN search filters[4] is an actively maintained collection of *"pre-tested strategies that identify the higher quality evidence from the vast amounts of literature indexed in the major medical databases."* It covers six study types: Systematic reviews, Randomised controlled trials, Observational studies, Diagnostic studies, Economic studies, and Patient issues. We also consulted the InterTASC Information Specialists' Sub-Group[5], who maintain a 'Search Filter Resource' as a *'collaborative venture to identify, assess and test search filters designed to retrieve research by study design or focus'*. On their advice [Glanville, personal communication], we augmented our collection with two further strategies on the topics of Diagnostic Studies and Economic Evaluations which had been the subject of expert reviews (J. Glanville, personal communication, November 1, 2017). This resulted in a total of eight actively maintained strategies, consisting of 47 disjunctions containing 355 terms (i.e. a mean of 7.55 terms per disjunction). Each term consists of a mean of 1.70 tokens.
3. A collection of recruitment search strategies. There is no standard test collection for recruitment search; in fact very little data of this type is made available publicly. However, there are various community initiatives to collect Boolean strings for recruitment, notably:

---

[3] Our test data is publicly available via https://github.com/tgr2uk/query_suggestions
[4] http://www.sign.ac.uk/search-filters.html
[5] https://sites.google.com/a/york.ac.uk/issg-search-filters-resource/

a. The Boolean Search Strings Repository[6]: a communal collection of recruitment search strings curated by Irina Shamaeva
b. The Boolean Search String Experiment[7]: a collection of Boolean strings collected by Glen Cathey to address a specific recruitment brief.

After deduplication, these two sources in combination yielded a total of 46 search strategies, containing 80 disjunctions with 571 terms (a mean of 7.15 terms per disjunction). Each term consists of a mean of 1.38 tokens.

In aggregate, these three sources represent data that is curated, actively maintained, and specific to more than one domain. In sum they contain a total of 74 expert search strategies consisting of 229 disjunctions and 1,824 individual query terms. To the best of our knowledge, our experiments represent the first study of this scale and coverage.

An example search strategy from the recruitment domain is shown in Figure 3 above. Examples from the CLEF data set and the SIGN data set are shown in Figures 4 and 5 respectively.

```
1  "Aspergillus"[MeSH]
2  "Aspergillosis"[MeSH]
3  "Pulmonary Aspergillosis"[MeSH]
4  aspergill*[tiab]
5  fungal infection[tw]
6  (invasive[tiab] AND fungal[tiab])
7  1 OR 2 OR 3 OR 4 OR 5 OR 6
8  "Serology"[MeSH]
9  Serology"[MeSH]
10 (serology[tiab] OR serodiagnosis[tiab] OR serologic[tiab])  11  8 OR 9 OR 10
12 "Immunoassay"[MeSH]
13 (immunoassay[tiab] OR immunoassays[tiab])
14 (immuno assay[tiab] OR immuno assays[tiab])
15 (ELISA[tiab] OR ELISAs[tiab] OR EIA[tiab] OR EIAs[tiab])
16 immunosorbent[tiab]
17 12 OR 13 OR 14 OR 15 OR 16
18 Platelia[tw]
19 "Mannans"[MeSH]
20 galactomannan[tw]
21 18 OR 19 OR 20
22 11 OR 17 OR 21
23 7 AND 22
```

**Fig 4**. An example search strategy from the CLEF 2017 data set

---

[6] https://booleanstrings.ning.com/forum/topics/boolean-search-strings-repository
[7] http://booleanblackbelt.com/2010/11/boolean-search-string-experiment-are-you-game/

```
1.  Meta-Analysis as Topic/
2.  meta analy$.tw.
3.  metaanaly$.tw.
4.  Meta-Analysis/
5.  (systematic adj (review$1 or overview$1)).tw.
6.  exp Review Literature as Topic/
7.  or/1-6
8.  cochrane.ab.
9.  embase.ab.
10. (psychlit or psyclit).ab.
11. (psychinfo or psycinfo).ab.
12. (cinahl or cinhal).ab.
13. science citation index.ab.
14. bids.ab.
15. cancerlit.ab.
16. or/8-15
17. reference list$.ab.
18. bibliograph$.ab.
19. hand-search$.ab.
20. relevant journals.ab.
21. manual search$.ab.
22. or/17-21
23. selection criteria.ab.
24. data extraction.ab.
25. 23 or 24
26. Review/
27. 25 and 26
28. Comment/
29. Letter/
30. Editorial/
31. animal/
32. human/
33. 31 not (31 and 32)
34. or/28-30,33
35. 7 or 16 or 22 or 27
36. 35 not 34
```

**Fig 5**. An example search strategy from the SIGN data set

### 2.3.1 Research questions

In this paper, we investigate the following research questions:

1. To what extent can methods based on manually curated ontologies provide suitable query suggestions for professional search?
2. To what extent can methods based on context-free distributional language models provide suitable query suggestions for professional search?

3. To what extent can combining the above methods improve on the performance of either method in isolation?

# 3. Materials and methods

As discussed above, in our experimental setup we investigate the extent to which different methods can predict gold standard data in the form of human-generated search strategies. We consider a variety of methods, as follows:

1. Noun phrases extracted from the result snippets of a commercial search engine (as a baseline)
2. Cluster labels generated by automatically clustering search result snippets
3. Related terms extracted from manually curated ontologies
4. Terms extracted from abstracts found within manually curated ontologies
5. Terms generated using context-free distributional language models trained on text corpora

We also investigate combining the above results in a variety of configurations.

## 3.1 Search result snippets

As a baseline, we hypothesize that the top matching results from a commercial search engine will provide a useful source of query suggestions, e.g. (Kruschwitz et al., 2009; Song et al., 2014). We extracted noun phrases from the top ten snippets returned by a popular Web search engine, in this case Google, restricting the search to domain-specific sites, e.g. PubMed[8] (for healthcare data) and Indeed[9] (for recruitment data). We identify phrases using the noun phrase extraction API of TextBlob[10] which in turn utilizes methods provided by the Natural Language Toolkit (Loper & Bird, 2002).

## 3.2 Cluster labels

Clustering tools may be used to generate query suggestions in the form of cluster labels generated from search result snippets. We used a popular, freely available clustering tool, Carrot2 (Stefanowski & Weiss, 2003), and configured it using the default settings for number of results and minimum cluster size, and then queried PubMed (for the healthcare data) and Wikipedia (for the recruitment data) to generate cluster labels using three different clustering algorithms (kMeans, Lingo and suffix tree clustering)[11]. Evidently, there is scope to customize this process further but our intent at this stage is to explore the underlying principle and provide a comparative baseline.

---

[8] https://www.ncbi.nlm.nih.gov/pubmed/
[9] https://www.indeed.com
[10] https://textblob.readthedocs.io/en/dev/ (using default settings)
[11] Carrot2 also offers additional search feeds through a partnership with the etools.ch metasearch engine, but these impose IP-based blocking and rate limiting that preclude systematic testing.

## 3.3 Ontological relations

Query suggestions can be generated by querying manually curated ontological resources to identify related terms in the form of hypernyms, hyponyms etc. Many such resources are hosted on the web as Linked Open Data[12], and support access via structured query languages such as SPARQL. Some are structured as formal ontologies (modelling subsumption and other relations), others as controlled vocabularies and thesauri. We investigated a variety of such resources, of both a general purpose and domain-specific nature. Given their wide coverage and generic nature, the first two resources may be considered general-purpose, and the latter four as domain-specific (to healthcare):

1. **DBpedia** is a project aiming to extract structured content from Wikipedia (Gangemi et al., 2018). The DBpedia data set describes 4.58 million entities, out of which 4.22 million are classified in a consistent ontology.
2. **WebISA** (Seitner et al., 2016) is a publicly available database containing hypernymy relations extracted from the CommonCrawl web corpus[13]. The LOD version contains 11.7 million hypernymy relations, each provided with rich provenance information and confidence estimates.
3. **Medical Subject Headings**[14] (MeSH) is a controlled vocabulary for the purpose of indexing documents in the life sciences. It contains a total of 25,186 *subject headings*, which are accompanied by a short description or definition, links to related descriptors, and a list of synonyms or very similar terms.
4. **RxNorm**[15] is a terminology that contains all medications available on the US market. It has concepts for drug ingredients, clinical drugs and dose forms.
5. The **British National Formulary** (BNF)[16] is a pharmaceutical reference that contains information about medicines available on the UK National Health Service (NHS).
6. **The DrugBank** database[17] is an online database containing information on drugs and drug targets. The latest release of DrugBank contains 11,683 drug entries, 1,117 approved biotech drugs, 128 nutraceuticals and over 5,505 experimental drugs.

We created SPARQL queries to their respective endpoints to retrieve related terms, and set the maximum number of results to the default of 100. In cases where querying a particular resource returned more than one type of related term (e.g. both 'broader' and 'narrower' terms), these were aggregated and returned as a single list.

---

[12] http://linkeddata.org/data-sets
[13] http://commoncrawl.org/
[14] https://www.nlm.nih.gov/mesh/meshhome.html
[15] https://www.nlm.nih.gov/research/umls/rxnorm/
[16] https://www.bnf.org/
[17] https://www.drugbank.ca/

## 3.4 Term extraction

In addition to providing related terms, some ontological resources include natural language descriptions for their entries. DBpedia, for example, provides abstracts that are written in an encyclopedic style. Likewise, MeSH contains definitions for many of its terms. These descriptions provide a further source of potential query suggestions. We created SPARQL queries to DBpedia and MeSH to retrieve the descriptions, then applied the following keyword extraction algorithms:

- Neoclassical combining forms (Díaz-Negrillo, 2014)
- TextBlob:nltk-np[18]
- Gensim:textrank[19] (Mihalcea & Tarau, 2004)
- Textacy:textrank[20]
- Textacy:sgrank[21] (Danesh et al., 2015)
- Rake-nltk[22] (Rose et al., 2010)

## 3.5 Context-free distributional language models

Word embeddings as a class of techniques where individual words or phrases are represented as real-valued dense vectors in a predefined vector space have become the de facto representation standard in many NLP applications (Jurafsky & Martin, 2020). Since they model the distributional patterns of words, they can be used to generate query suggestions in the form of related terms. Word embeddings can be learned from text corpora using a variety of techniques, e.g. word2vec (T Mikolov et al., 2013), GloVe (Pennington et al., 2014), FastText (Bojanowski et al., 2017), BERT (Devlin et al., 2019) etc. A number of publicly available, pre-built embedding models are available, trained on sources such as Wikipedia (Pennington et al., 2014), GoogleNews (Tomas Mikolov et al., 2013), and PubMed (Chiu et al., 2016). Given that our evaluation approaches considers query terms *in isolation*, we do not deploy *contextual* embeddings (such as BERT) but investigate the following *context-free* embeddings:

- Word2vec trained on Google news (Tomas Mikolov et al., 2013)
- GloVe trained on Wikipedia + Gigaword5 (Pennington et al., 2014)
- FastText trained on Wikipedia (Bojanowski et al., 2017)
- Word2vec trained on PubMed articles, with different window sizes (2 and 30) (Chiu et al., 2016)

---

[18] https://textblob.readthedocs.io/en/dev
[19] https://radimrehurek.com/gensim/summarization/summariser.html
[20] https://chartbeat-labs.github.io/textacy/api_reference.html?highlight=textrank#textacy.keyterms.textrank
[21] https://chartbeat-labs.github.io/textacy/api_reference.html?highlight=sgrank#textacy.keyterms.sgrank
[22] https://pypi.org/project/rake-nltk/

We also built bespoke models using the PubMed Open Access full text snapshot from September 2017, which consisted of 944,672 full-text articles. Using an initial test set we identified the optimal parameter settings as dimensions=300, window size=5, min word count=10. We created two bespoke Word2vec models: one which consisted solely of unigrams, and a second model which also included bigrams and trigrams.

# 4. Results

Our overall evaluation approach was as follows: for every strategy in our test collection, we iterate over each disjunction and calculate precision, recall and F score for each term, based on the overlap between the suggested term set and the gold standard. Although search strategies may also contain conjunctions and other expressions, they are in general not a useful part of the gold standard data as they do not represent sets of synonyms or closely related terms. We then repeat this process for each method, and report performance in terms of average (arithmetic mean of) precision, recall and F score[23]. We test for significance using one-way ANOVA, and report values where $p < 0.01$.

## 4.1 Search result snippets

Table 1 shows the arithmetic mean of precision (P) and recall (R) and the F score (F) for the noun phrases extracted from Google snippets, with the highest F value highlighted in bold.

| CLEF 2017 (n=898) | | | SIGN (n=355) | | | Recruitment (n=571) | | |
|---|---|---|---|---|---|---|---|---|
| P | R | F | P | R | F | P | R | F |
| 0.015 | 0.021 | 0.018 | 0.009 | 0.011 | 0.010 | 0.023 | 0.036 | **0.028** |

Table 1: Precision, recall and F for Google snippets

Although these figures may appear low in absolute terms, they are in line with the findings of similar studies applying the same methodology to digital libraries (Kruschwitz et al., 2009) and local websites and intranets (Adeyanju et al., 2012). This reflects the difficulty in predicting query suggestions based on a ground truth of nothing more than terms found in existing disjunctions. Moreover, they represent a likely underestimate of performance, since some of the terms identified as false positives may transpire to be acceptable in a real task scenario (see Discussion).

---

[23] Standard deviation values have been omitted for reasons of brevity

## 4.2 Cluster labels

Carrot2[24] supports three clustering algorithms: Lingo, suffix tree clustering (STC), and kMeans. The means of P, R and F for these three algorithms are shown in Table 2, with the highest F value in each column highlighted in bold (as in all the following tables).

|  | **CLEF 2017 (n=898)** | | | **SIGN (n=355)** | | | **Recruitment (n=571)** | | |
|---|---|---|---|---|---|---|---|---|---|
|  | P | R | F | P | R | F | P | R | F |
| **Lingo** | 0.005 | 0.009 | 0.006 | 0.002 | 0.004 | 0.003 | 0.008 | 0.019 | 0.011 |
| **STC** | 0.034 | 0.042 | **0.038** | 0.017 | 0.027 | **0.021** | 0.035 | 0.089 | **0.050** |
| **kMeans** | 0.015 | 0.025 | 0.019 | 0.003 | 0.008 | 0.004 | 0.014 | 0.025 | 0.018 |

Table 2: Precision, recall and F for Carrot2 cluster labels

Overall, STC performs best, with F values ranging from 0.05 (Recruitment) to 0.021 (SIGN). kMeans is consistently in second place and Lingo third. Comparing F scores shows that the choice of clustering algorithm has a significant effect on performance for CLEF, $F(2, 2691) = 112.38$, $p < 0.01$, for SIGN $F(2, 1062) = 54.05$, $p < 0.01$ and for Recruitment $F(2, 1710) = 79.35$, $p < 0.01$. Interestingly, this result runs contrary to the findings of Carrotsearch's own evaluation of cluster label quality[25], but this may reflect the difference between generating cluster labels for human interpretation vs. predicting related terms found in search strategies. Using STC clearly outperforms the previous baseline.

## 4.3 Ontological relations

|  | **CLEF 2017 (n=898)** | | | **SIGN (n=355)** | | | **Recruitment (n=571)** | | |
|---|---|---|---|---|---|---|---|---|---|
|  | P | R | F | P | R | F | P | R | F |
| DBpedia | 0.026 | 0.046 | **0.033** | 0.024 | 0.034 | **0.028** | 0.019 | 0.043 | **0.026** |
| WebISA | 0.013 | 0.010 | 0.011 | 0.014 | 0.009 | 0.011 | 0.005 | 0.004 | 0.004 |
| MeSH | 0.065 | 0.017 | 0.027 | 0.148 | 0.015 | 0.027 | n/a | n/a | n/a |
| RxNorm | 0.000 | 0.000 | 0.000 | 0.000 | 0.000 | 0.000 | n/a | n/a | n/a |
| BNF | 0.002 | 0.001 | 0.001 | 0.000 | 0.000 | 0.000 | n/a | n/a | n/a |
| DrugBank | 0.000 | 0.000 | 0.000 | 0.000 | 0.000 | 0.000 | n/a | n/a | n/a |

Table 3: Precision, recall and F for manually curated resources

---

[24] https://project.carrot2.org/algorithms.html
[25] https://carrotsearch.com/lingo3g/comparison/

The results for the manually curated resources are shown in Table 3. Overall, these results are comparable with those of the Carrot2 cluster labels, with the highest F score being 0.033. Comparing F scores for the general purpose resources (DBpedia vs. WEBISA) shows a significant difference in favour of the former on all three data sets, particularly Recruitment $F(1, 1140) = 59.20$, $p < 0.01$.

Comparing the F scores across all sources shows that the source of suggested terms has a significant effect on performance for both CLEF, $F(5, 5382) = 109.53$, $p < 0.01$ and SIGN $F(5, 2124) = 62.03$, $p < 0.01$. The use of a specialist resource appears to be beneficial in terms of precision, with relatively high values shown by MeSH (0.148 for SIGN data). This reflects the highly specialised nature of this resource. However, the best performing resource overall (in terms of F measure) remains DBpedia.

## 4.4 Term extraction

|  | CLEF 2017 (n=898) | | | SIGN (n=355) | | | Recruitment (n=571) | | |
|---|---|---|---|---|---|---|---|---|---|
|  | P | R | F | P | R | F | P | R | F |
| NCF regex | 0.026 | 0.030 | 0.028 | 0.010 | 0.011 | 0.010 | 0.012 | 0.023 | 0.016 |
| Nltk-np | 0.017 | 0.017 | 0.017 | 0.011 | 0.011 | 0.011 | 0.018 | 0.030 | 0.023 |
| rake | 0.005 | 0.004 | 0.004 | 0.004 | 0.004 | 0.004 | 0.003 | 0.005 | 0.004 |
| sgrank | 0.024 | 0.023 | 0.023 | 0.013 | 0.014 | **0.013** | 0.008 | 0.011 | 0.009 |
| Gensim textrank | 0.027 | 0.017 | 0.021 | 0.010 | 0.009 | 0.009 | 0.016 | 0.014 | 0.015 |
| Textacy textrank | 0.030 | 0.033 | **0.031** | 0.012 | 0.014 | **0.013** | 0.020 | 0.024 | **0.022** |

Table 4: Precision, recall and F for terms extracted from DBpedia abstracts

The results for keywords extracted from DBpedia abstracts are shown in Table 4. Overall, the scores are slightly lower than those of the manually curated terms. Comparing F scores shows that the keyword extraction algorithm has a significant effect on performance, with Textacy textrank returning the highest F measure (or equal highest) across all datasets: CLEF $F(5, 5382) = 28.19$, $p < 0.01$, SIGN $F(5, 2124) = 3.51$, $p < 0.01$ and Recruitment $F(5, 3420) = 16.89$, $p < 0.01$.

|  | CLEF 2017 (n=898) | | | SIGN (n=355) | | |
|---|---|---|---|---|---|---|
|  | P | R | F | P | R | F |
| NCF regex | 0.018 | 0.011 | **0.014** | 0.013 | 0.004 | **0.006** |
| Nltk-np | 0.011 | 0.006 | 0.008 | 0.010 | 0.005 | **0.006** |
| rake | 0.007 | 0.005 | 0.006 | 0.003 | 0.003 | 0.003 |
| sgrank | 0.013 | 0.010 | 0.011 | 0.006 | 0.006 | **0.006** |
| Gensim textrank | 0.003 | 0.001 | 0.002 | 0.004 | 0.001 | 0.002 |
| Textacy textrank | 0.014 | 0.013 | 0.013 | 0.004 | 0.006 | 0.005 |

Table 5: Precision, recall and F for terms extracted from MeSH descriptions

The results for keyword extraction applied to MeSH descriptions (using healthcare data) are shown in Table 5. In almost all cases, these scores are lower than the equivalent returned by DBpedia abstracts. Comparing F scores shows that the keyword extraction algorithm has a significant effect on performance for CLEF, with NCF regex performing best $F(5, 5382) = 10.46$, $p < 0.01$. It also performs joint best on SIGN, although this effect is not significant.

## 4.5 Context-free distributional language models

|  | CLEF 2017 (n=898) | | | SIGN (n=355) | | | Recruitment (n=571) | | |
|---|---|---|---|---|---|---|---|---|---|
|  | P | R | F | P | R | F | P | R | F |
| Word2vec+Google News | 0.033 | 0.037 | 0.035 | 0.027 | 0.025 | 0.028 | 0.041 | 0.035 | 0.038 |
| GloVe+Wikipedia | 0.044 | 0.047 | 0.045 | 0.026 | 0.030 | 0.028 | 0.057 | 0.047 | **0.051** |
| FastText+Wikipedia | 0.024 | 0.038 | 0.029 | 0.019 | 0.016 | 0.017 | 0.024 | 0.018 | 0.021 |
| Word2vec+PubMed (win2) | 0.057 | 0.062 | 0.059 | 0.026 | 0.028 | 0.027 | n/a | n/a | n/a |
| Word2vec+PubMed (win30) | 0.069 | 0.073 | 0.071 | 0.028 | 0.033 | 0.030 | n/a | n/a | n/a |
| Bespoke word2vec +PubMed, unigrams | 0.071 | 0.075 | **0.073** | 0.038 | 0.040 | 0.039 | n/a | n/a | n/a |
| Bespoke word2vec +PubMed, trigrams | 0.069 | 0.072 | 0.072 | 0.042 | 0.040 | **0.041** | n/a | n/a | n/a |

Table 6: Precision, recall and F for distributional models

The results for the language models are shown in Table 6. Overall, these scores are generally higher than those of previous methods. Comparing F scores shows that the choice of model has a significant effect on performance, although the pattern is inconsistent: the bespoke PubMed unigram model performs best on CLEF $F(6, 6279) = 27.49$, $p < 0.01$, while the bespoke PubMed trigram model performs the best on SIGN $F(6, 2478) = 6.19$, $p < 0.01$. Their performance is comparable to that of Word2vec+PubMed (win30) (Chiu et al., 2016), which provides some evidence for the reproducibility of these results. Comparing the three generic models on recruitment data, GloVe+Wikipedia performs best $F(2, 1710) = 19.78$, $p < 0.01$.

These results illustrate the value of using domain-specific models (the lower half of the table) rather than generic models (the upper half). The fact that the two bespoke models outperformed the pre-trained models is also interesting (although for CLEF this difference is not significant). One possible explanation may be that the bespoke models were created using a relatively clean corpus which included only body text (i.e. no figures, headers, footers, etc.) and excluded numbers, punctuation and non-alphabetic characters.

## 4.6 Combining sources

A primary motivation for the work in this paper is to facilitate the development of practical applications (as opposed to adopting a purely academic perspective). With this in mind, the following section explores how to make optimal use of different resources in a variety of combinations.

For example, it may be possible to improve performance (particularly in terms of recall) by combining results from two or more sources. Evidently, the nature of that improvement will depend on the particular services being combined and the way in which their respective result sets intersect. Not only does this present an interesting theoretical question, but it also offers the prospect of significant impact on a large proportion of the professional search community. In this section we investigate the effects of combining the best performing curated resources with the best performing language models.

### 4.6.1 Simple aggregation

The simplest form of aggregation is to combine two term suggestion sets as a 'bag of words' (note that since the evaluation is based on set overlap their order is not significant). Table 7 shows the results of applying a combination of the DBpedia ontology and the GloVe+Wikipedia language model to recruitment data (also showing the results for each method in isolation).

In this instance, combining two sources improves recall, but at the expense of precision, with a decrease in F score (compared to GloVe in isolation). Comparing F scores shows that aggregation has a significant effect on performance $F(2, 1710) = 20.14$, $p < 0.01$.

|  | Recruitment (n=571) | | |
| --- | --- | --- | --- |
|  | P | R | F |
| DBpedia (alone) | 0.019 | 0.043 | 0.026 |
| GloVe+Wikipedia (alone) | **0.057** | 0.047 | **0.051** |
| Aggregated | 0.030 | **0.081** | 0.044 |

Table 7: Precision, recall and F for simple aggregation of terms from DBPEDIA and GloVe

Table 8 shows the results of combining the MeSH ontology with the word2vec PubMed trigram language model for healthcare (also showing the results for each method in isolation). In this instance, the combination offers improvements in both recall and F score for both data sets. Comparing F scores shows that the use of aggregation has a consistently positive and significant effect on performance on both CLEF $F(2, 2691) = 78.57$, $p < 0.01$ and SIGN $F(2, 1062) = 5.36$, $p < 0.01$.

|  | CLEF 2017 (n=898) | | | SIGN (n=355) | | |
| --- | --- | --- | --- | --- | --- | --- |
|  | P | R | F | P | R | F |
| MeSH (alone) | **0.065** | 0.017 | 0.027 | **0.148** | 0.015 | 0.027 |
| Bespoke PubMed trigram (alone) | 0.071 | 0.075 | 0.073 | 0.042 | **0.040** | 0.041 |
| Aggregated | 0.082 | **0.081** | **0.081** | 0.075 | 0.035 | **0.048** |

Table 8: Precision, recall and F for simple aggregation of terms from MeSH and PubMed trigram model

### 4.6.2 Back-off approaches

One possible explanation for the positive effect of aggregation is that language models tend to learn robust representations for frequent terms, which tends to favour unigrams. By contrast, manually curated ontologies tend to provide better coverage of higher order ngrams (bigrams and above), which reflects their focus on named entities and other specialist terminology. To test this hypothesis, we implemented two further combinations which exploited the ngram order in finding related terms. Both these approaches represent *back-off* algorithms of the sort that has long been popular in a variety of NLP applications in cases where data sparsity has been an issue (Manning et al., 2008):

**Agg2: 'Loose pipelining':**
1. Tokenize the query term (based on whitespace)
2. If number of tokens > 1, look up term (ngram) in curated ontology

3. Look up term (unigram or ngram) in language model
4. Combine results and return as a unified list

**Agg3: 'Strict pipelining':**
1. Tokenize the query term (based on whitespace)
2. If number of tokens > 1, look up term (ngram) in curated ontology
    a. If no results from curated ontology, look up term (ngram) in language model
3. Else look up term (unigram) in language model
4. Combine results and return as a unified list

What these approaches have in common is that curated resources are only used for higher order ngrams (bigrams and above). Where they differ is that in the second variation the language model is only used if the curated ontology returned no results or if the term is a unigram. Table 9 shows the results of this approach, along with the results from the approaches above (repeated here for convenience): the best performing curated ontology (MeSH for healthcare, and DBpedia for recruitment); the best performing language model (PubMed trigram for healthcare, GloVe for recruitment), and simple aggregation (shown here as 'Agg1'). The lower two rows show the results for 'loose pipelining' (Agg2) and 'strict pipelining' (Agg3).

|  | CLEF 2017 (n=898) | | | SIGN (n=355) | | | Recruitment (n=571) | | |
| --- | --- | --- | --- | --- | --- | --- | --- | --- | --- |
|  | P | R | F | P | R | F | P | R | F |
| Curated ontology | 0.065 | 0.017 | 0.027 | 0.148 | 0.015 | 0.027 | 0.019 | 0.043 | 0.026 |
| Language model | 0.071 | 0.075 | 0.073 | 0.042 | 0.040 | 0.041 | 0.057 | 0.047 | 0.051 |
| Agg1 | 0.082 | **0.081** | 0.081 | 0.073 | **0.074** | 0.035 | 0.030 | **0.081** | 0.044 |
| Agg2 | 0.083 | 0.081 | 0.082 | 0.075 | 0.035 | 0.048 | 0.061 | 0.069 | 0.065 |
| Agg3 | **0.100** | 0.076 | **0.086** | **0.135** | 0.032 | **0.052** | **0.065** | 0.068 | **0.066** |

Table 9: Precision, recall and F for combinations using backoff approaches

The results show that simple aggregation (Agg1) consistently produces the highest recall, which reflects the undifferentiated, broader nature of a combined suggested terms list. Conversely, 'strict pipelining' (Agg3) consistently produces the highest precision, which supports the hypothesis that ngram order can be exploited when finding related terms. Moreover, the F scores show that it is possible to combine suggestions from different sources using strict pipelining to deliver a more effective balance of precision & recall.

# 5. Discussion

We will approach the discussion from a number of different angles representing different variables in our experimental setup. First of all we frame the discussion by reviewing some of the key assumptions behind this type of study and how it differs from prior studies. It is important to recognise that although the use of query expansion has been the subject of many studies, relatively few have focused explicitly on the professional search context. Moreover, the few that have done so are generally predicated on the assumption that users will adopt a simplistic approach based on unstructured keyword queries, e.g. (Lu et al., 2009). To the best of our knowledge this is the first study of this scale to evaluate interactive expansion within the context of structured queries using publicly available, human-generated search strategies[26].

Turning to the results themselves, we may make a few general observations. First, although some of the results may appear low in absolute terms (e.g. a maximum F-score of 0.086), the key observation is that relative differences are statistically significant and generalisable. Moreover, despite the ostensibly modest absolute values, the potential impact on professional search practice could be significant: with patent search tasks taking a median of 12 hours to complete (Russell-Rose et al., 2018), even a 10% saving due to improved query formulation would translate to 1.2 hours of billable time per task. Likewise, librarians spend an average aggregated time of 26.9 hours on systematic reviews, most of which is spent on search strategy development and translation (Bullers et al., 2018). Query expansion is known to be highly valued by healthcare information professionals, so the potential for adoption of even imperfect query suggestion techniques could lead to considerable impact.

Comparing the different techniques, we see that the use of language models outperforms methods based on manually-curated resources. This includes both ontological relations and terms extracted from abstracts or definitions. It is possible of course that other human-curated resources may offer improved performance, e.g. ConceptNet[27], Wikidata[28], etc. However, the six sources investigated in this study offer a reasonable basis for comparison, and the investigation of additional resources is suggested as an area for further work.

In addition to the above, the practice of combining sources offers the prospect of further improvement, with simple aggregation having a consistently positive and significant effect on recall across all data sets. Moreover, it is possible to deliver a better balance between precision & recall by utilizing ngram order in the combination, e.g. using strict pipelining to optimise for precision.

It is important also to recognise that the results represent a lower bound on potential performance, since some of the terms identified as false positives may transpire to be true positives in a live task scenario. For example, the first disjunction in the recruitment data set contains the terms:

---

[26] All test data is publicly available via https://github.com/tgr2uk/query_suggestions
[27] http://conceptnet.io/
[28] https://www.wikidata.org/wiki/Wikidata:Main_Page

```
['analyst', 'business analyst', 'business process analyst',
'data analyst', 'reporting analyst']
```

When DBPEDIA is queried using the second of these terms ('business analyst'), it returns the following suggestions:

```
['BA', 'Business occupations', 'Business terms', 'Systems
analysis', 'Functional analyst', 'Software Business Analyst',
'Business analysis', 'Computer occupations', 'Business systems
analyst', 'Analyst']
```

Arguably, the terms 'BA', 'Software business analyst', 'Business systems analyst' and 'Analyst' are all true positives. However, due to the offline evaluation process they are all labelled as false positives apart from 'Analyst', resulting in a precision of 0.1 instead of 0.4. Moreover, had the term 'BA' (a common abbreviation for 'business analyst') been included in the original disjunction, the recall would be 0.333 instead of 0.2.

This observation brings us naturally onto the limitations of this study. Although the test data represents a sizable collection of search strategies, there is no guarantee that they are optimal, i.e. they represent an 'ideal' articulation of the information needs they represent. Indeed, the very fact that they were created without access to the type of query formulation techniques proposed in this paper would imply that they are less than 'perfect'. However, this does not mean they are without value: the majority are drawn from hand-curated, published and publicly maintained sources, and represent the work of trained experts. They may not be ideal, but they are representative of a broader population, and in this respect we believe they are a valid approximation of professional search behaviour.

Evidently, to accurately evaluate how real users would react in a real task scenario, it is necessary to set up a user study involving representative human participants. This is of course more expensive and time consuming, and user studies can be more challenging to scale and replicate. In this respect the value of this study is in investigating a diverse set of techniques using human generated search strategies as a proxy for human behaviour. As such it offers a scalable and reproducible approach which allows more expensive online studies to be better focused on specific issues and tasks.

A further limitation of this study is that we have treated disjunctions ('OR' clauses) in the data as the primary unit of analysis. Evidently, search strategies contain other types of construction (e.g. conjunctions, negations, etc.) and these may offer additional evaluation possibilities. Finally, our use of live, publicly available LOD endpoints facilitates transparency and reproducibility, but at the expense of occasional latency issues and timeouts. To mitigate this issue, all runs were replicated at least once to ensure consistency and reproducibility.

# 6. Conclusions and further work

In this paper, we review the role of query suggestions within the context of professional search strategies used in real-world expert search tasks. We investigate a number of techniques for generating query suggestions, and evaluate them using a variety of data sources. We now draw conclusions in relation to the original research questions set out in Section 3.

1. *To what extent can methods based on manually curated ontologies provide suitable query suggestions for professional search?*

   We found that the ontological relations in generic, manually curated resources such DBpedia outperformed the baseline of search results snippets for healthcare search strategies but not for recruitment search strategies. Even when using domain-specific resources, the performance was poorer than that of extracting cluster labels from the search results snippets.

   The use of terms extracted from abstracts and definitions was not shown to be effective. When using generic resources (e.g. DBPEDIA), the results were an improvement over the baseline for healthcare but not for recruitment. Terms extracted from domain-specific resources consistently performed worse than the baseline.

2. *To what extent can methods based on context-free distributional language models provide suitable query suggestions for professional search?*

   We found that context-free distributional language models outperformed the baseline for all data sets. They also outperformed the use of manually-curated resources (whether used as a source of ontological relations or as a source of terms in abstracts/definitions). We also found that our own bespoke Pubmed model outperformed the best of the 3rd party pre-built models on healthcare data. The best performing model on recruitment data was found to be GloVe+Wikipedia.

3. *To what extent can combining the above methods improve on the performance of either method in isolation?*

   We found that simple aggregation consistently produced higher recall than any method in isolation. This gave rise to a higher F score for both the healthcare data sets, but not for recruitment data, where the highest average F score continued to be that of GloVe+Wikipedia (alone).

The use of aggregate methods showed that it is possible to exploit ngram order in finding related terms. 'Strict pipelining' consistently produced the highest precision and highest overall F score, which demonstrates that it is possible to combine suggestions from different sources to deliver a better overall balance of precision & recall.

## 6.1 Future work

This work provides a benchmark set of results (in an under-explored area) for future experiments. A valuable next step would be to scale the work horizontally, e.g. to other curated resources (such as ConceptNet[29] and Wikidata[30]) or to other distributional models and frameworks. The NLP field is actively growing and new distributional approaches are continually being developed, and it may also be productive to explore other bespoke models, e.g. for recruitment data. Given the effectiveness of the context-free embeddings in our experiments and the impact of contextualised embeddings such as BERT (Devlin et al., 2019) across a variety of NLP tasks, a further next step may be to explore contextual embeddings, for example using neighboring disjunction terms as context.

A further form of scaling is to investigate other domains: in this study we focused on healthcare and recruitment, aligning with two professions known to be among the heaviest users of complex, Boolean queries. It would be interesting to extend this work to other professions such as patent search, competitive intelligence, and media monitoring (Russell-Rose et al., 2018).

Another possibility is to revisit the test data and explore constructs other than disjunctions (e.g. operators such AND, ADJ, etc.). These were deemed out of scope due to their inconsistent semantics, but it is possible that other consistent types of relation may be identified which may form an additional focus for evaluation.

Finally, a further area for future work is to compare these findings with human judgements as might be elicited via a user study. This work could explore the degree to which our findings align with that of naturalistic use, and determine the extent to which false positives identified in our study may actually transpire to be true positives in live, interactive usage.

---

[29] http://conceptnet.io/
[30] https://www.wikidata.org/wiki/Wikidata:Main_Page

# Declarations

## Funding

This research was supported by Innovate UK Open Competition R&D grant 102975, "Intelligent Search Assistance". Innovate UK had no involvement in the study design, data analysis, report writing or decision to submit for publication.

## Declaration of competing interest

The authors declare that they have no known competing financial interests or personal relationships that could have appeared to influence the work reported in this paper.

## Availability of data and material

The datasets used in this paper were acquired and curated from publicly available resources (see Section 2.3).

## Code availability

Test data is publicly available via Github. Evaluation code is hosted on BitBucket and can be made available on demand.

## CRediT authorship contribution

Tony Russell-Rose: Funding acquisition, Resources, Conceptualization, Methodology, Software, Data curation, Project administration, Writing - original draft.
Phil Gooch: Resources, Conceptualization, Methodology, Resources, Software, Data curation, Writing - review & editing.
Udo Kruschwitz: Conceptualization, Methodology, Writing - review & editing.

Bibliography

Adeyanju, I. A., Song, D., Albakour, M., Kruschwitz, U., De Roeck, A., & Fasli, M. (2012).

    Adaptation of the concept hierarchy model with search logs for query recommendation

    on intranets. *SIGIR*, 5–14.

Aggarwal, N., & Buitelaar, P. (2012). *Query Expansion Using Wikipedia and Dbpedia*. CLEF

(Online Working Notes/Labs/Workshop).

Albakour, M.-D., Kruschwitz, U., Nanas, N., Kim, Y., Song, D., Fasli, M., & De Roeck, A. (2011). Autoeval: An evaluation methodology for evaluating query suggestions using query logs. *ECIR*, 605–610.

Balaneshinkordan, S., & Kotov, A. (2019). Bayesian approach to incorporating different types of biomedical knowledge bases into information retrieval systems for clinical decision support in precision medicine. *Journal of Biomedical Informatics*, *98*, 103238. https://doi.org/10.1016/j.jbi.2019.103238

Baron, J. R., Lewis, D. D., & Oard, D. W. (2006). TREC 2006 Legal Track Overview. *TREC*.

Bastian, H., Glasziou, P., & Chalmers, I. (2010). Seventy-five trials and eleven systematic reviews a day: how will we ever keep up? *PLoS Medicine*, *7*(9), e1000326. https://doi.org/10.1371/journal.pmed.1000326

Beitzel, S. M., Jensen, E. C., Chowdhury, A., Frieder, O., & Grossman, D. (2007). Temporal analysis of a very large topically categorized web query log. *Journal of the American Society for Information Science and Technology*, *58*(2), 166–178.

Bhogal, J., Macfarlane, A., & Smith, P. (2007). A review of ontology based query expansion. *Information Processing & Management*, *43*(4), 866–886. https://doi.org/10.1016/j.ipm.2006.09.003

Bojanowski, P., Grave, E., Joulin, A., & Mikolov, T. (2017). Enriching word vectors with subword information. *Transactions of the ACL*, *5*, 135–146.

Broder, A. (2002). A taxonomy of web search. *ACM SIGIR Forum*, *36*(2), 3. https://doi.org/10.1145/792550.792552

Bullers, K., Howard, A. M., Hanson, A., Kearns, W. D., Orriola, J. J., Polo, R. L., & Sakmar, K. A. (2018). It takes longer than you think: librarian time spent on systematic review tasks.

Journal of the Medical Library Association: JMLA, 106(2), 198.

Cao, G., Nie, J.-Y., Gao, J., & Robertson, S. (2008). Selecting good expansion terms for pseudo-relevance feedback. *SIGIR*, 243. https://doi.org/10.1145/1390334.1390377

Chiu, B., Crichton, G., Korhonen, A., & Pyysalo, S. (2016). How to Train good Word Embeddings for Biomedical NLP. *Proceedings of the 15th Workshop on Biomedical NLP*, 166–174. https://doi.org/10.18653/v1/W16-2922

Collobert, R., Weston, J., Bottou, L., Karlen, M., Kavukcuoglu, K., & Kuksa, P. (2011). Natural Language Processing (Almost) from Scratch. *The Journal of Machine Learning Research*, *12*, 2493–2537.

Danesh, S., Sumner, T., & Martin, J. H. (2015). Sgrank: combining statistical and graphical methods to improve the state of the art in unsupervised keyphrase extraction. *Proceedings of the Fourth Joint Conference on Lexical and Computational Semantics*, 117–126. https://doi.org/10.18653/v1/S15-1013

Devlin, J., Chang, M., Lee, K., & Toutanova, K. (2019). BERT: Pre-training of Deep Bidirectional Transformers for Language Understanding. *Proceedings of NAACL*, 4171–4186.

Díaz-Negrillo, A. (2014). Neoclassical compounds and final combining forms in English. *Linguistik Online*, *68*(6), 3–21.

Diaz, F., Mitra, B., & Craswell, N. (2016). Query Expansion with Locally-Trained Word Embeddings. *Proceedings of the 54th Annual Meeting of the ACL*, 367–377.

Efthimiadis, E. (1996). Query Expansion. *Annual Review of Information Science and Technology (ARIST)*, *31*, 121–187.

Elsweiler, D., Wilson, M. L., & Harvey, M. (2012). Searching for fun: Casual-leisure search. *ECIR 2012 Workshops*, *875*.

Gangemi, A., Navigli, R., Vidal, M.-E., Hitzler, P., Troncy, R., Hollink, L., Tordai, A., & Alam, M.

(Eds.). (2018). *The Semantic Web* (Vol. 10843). Springer. https://doi.org/10.1007/978-3-319-93417-4

Gibbs, A. (2006). *Heuristic Boolean patent search: comparative patent search quality/cost evaluation super Boolean vs. legacy Boolean search engines* [Technical Report]. Patent cafe.

Glanville, J. (2017, November 1). *Glanville, personal communication* [Personal communication].

Goeuriot, L., Kelly, L., Suominen, H., Névéol, A., Robert, A., Kanoulas, E., Spijker, R., Palotti, J., & Zuccon, G. (2017). *CLEF 2017 eHealth evaluation lab overview*. 291–303.

Gooda Sahib, N., Tombros, A., & Ruthven, I. (2010). Enabling Interactive Query Expansion through Eliciting the Potential Effect of Expansion Terms. In C. Gurrin, Y. He, G. Kazai, U. Kruschwitz, S. Little, T. Roelleke, S. Rüger, & K. van Rijsbergen (Eds.), *Advances in information retrieval* (Vol. 5993, pp. 532–543). Springer Berlin Heidelberg. https://doi.org/10.1007/978-3-642-12275-0_46

Griffon, N., Chebil, W., Rollin, L., Kerdelhue, G., Thirion, B., Gehanno, J.-F., & Darmoni, S. J. (2012). Performance evaluation of Unified Medical Language System®'s synonyms expansion to query PubMed. *BMC Medical Informatics and Decision Making*, *12*, 12. https://doi.org/10.1186/1472-6947-12-12

Grossman, M. R., Cormack, G. V., & Roegiest, A. (2016). TREC 2016 Total Recall Track Overview. *TREC*.

Hersh, W., Turpin, A., Price, S., Kraemer, D., Olson, D., Chan, B., & Sacherek, L. (2001). Challenging conventional assumptions of automated information retrieval with real users: Boolean searching and batch retrieval evaluations. *Information Processing & Management*, *37*(3), 383–402.

Jurafsky, D., & Martin, J. (2020). *Speech and Language Processing: An Introduction to Natural Language Processing,Computational Linguistics, and Speech Recognition* (third (draft)).

Karlgren, J. (2019). Adopting Systematic Evaluation Benchmarks in Operational Settings. In *Information Retrieval Evaluation in a Changing World: Lessons Learned from 20 Years of CLEF* (pp. 583–590). Springer International Publishing.

Kelly, D., Gyllstrom, K., & Bailey, E. W. (2009). *A comparison of query and term suggestion features for interactive searching*. 371–378.

Kim, Y., Seo, J., & Croft, W. B. (2011). Automatic boolean query suggestion for professional search. *Proceedings of the 34th International ACM SIGIR Conference*, 825–834.

Koster, C. H., Oostdijk, N. H., Verberne, S., & D'hondt, E. K. (2009). Challenges in professional search with phasar. *Proceedings of the Dutch-Belgian Information Retrieval Workshop*, 101–102.

Kruschwitz, U., Albakour, M.-D., Niu, J., Leveling, J., Nanas, N., Kim, Y., Song, D., Fasli, M., & De Roeck, A. (2009). Moving towards adaptive search in digital libraries. In *Advanced Language Technologies for Digital Libraries* (pp. 41–60). Springer.

Kruschwitz, U., Lungley, D., Albakour, M., & Song, D. (2013). Deriving query suggestions for site search. *Journal of the American Society for Information Science and Technology*, *64*(10), 1975–1994.

Kumar, A., Dandapat, S., & Chordia, S. (2020). Translating web search queries into natural language questions. *ArXiv Preprint ArXiv:2002.02631*.

Kuzi, S., Shtok, A., & Kurland, O. (2016). Query expansion using word embeddings. *Proceedings of CIKM '16*, 1929–1932. https://doi.org/10.1145/2983323.2983876

Lin, J. (2019). The Neural Hype, Justified! A Recantation. *SIGIR Forum*, *53*(2), 88–93.

Liu, Y., Miao, J., Zhang, M., Ma, S., & Ru, L. (2011). How do users describe their information need: Query recommendation based on snippet click model. *Expert Systems with Applications*, *38*(11), 13847–13856.

Loper, E., & Bird, S. (2002). NLTK: the natural language toolkit. *ArXiv Preprint Cs/0205028*.

Lupu, M., Mayer, K., Kando, N., & Trippe, A. J. (Eds.). (2011). *Current challenges in patent information retrieval* (Vol. 37). Springer. https://doi.org/10.1007/978-3-662-53817-3

Lu, Z., Wilbur, W. J., McEntyre, J. R., Iskhakov, A., & Szilagyi, L. (2009). Finding query suggestions for PubMed. *AMIA Annual Symposium Proceedings*, *2009*, 396–400.

Manning, C. D., Raghavan, Prabhakar., & SchützE, Hinrich. (2008). *Introduction to information retrieval*. Cambridge University Press.

Mihalcea, R., & Tarau, P. (2004). TextRank: Bringing Order into Texts. *EMNLP*, 404–411.

Mikolov, T, Sutskever, I., Chen, K., Corrado, G. S., & Dean, J. (2013). Distributed representations of words and phrases and their compositionality. *Advances in Neural Information Processing Systems*, 3111.

Mikolov, Tomas, Chen, K., Corrado, G., & Dean, J. (2013). Efficient estimation of word representations in vector space. *ArXiv Preprint ArXiv:1301.3781*.

Mitra, B., & Craswell, N. (2018). An introduction to neural information retrieval. *Foundations and Trends® in Information Retrieval*, *13*(1), 1–126. https://doi.org/10.1561/1500000061

Mullins, M. M., DeLuca, J. B., Crepaz, N., & Lyles, C. M. (2014). Reporting quality of search methods in systematic reviews of HIV behavioral interventions (2000-2010): are the searches clearly explained, systematic and reproducible? *Research Synthesis Methods*, *5*(2), 116–130. https://doi.org/10.1002/jrsm.1098

Navigli, R., & Velardi, P. (2003). An Analysis of Ontology-based Query Expansion Strategies. *ECML*. ECML.

Pennington, J., Socher, R., & Manning, C. (2014). Glove: global vectors for word representation. *Proceedings of the 2014 Conference on Empirical Methods in Natural Language Processing (EMNLP)*, 1532–1543. https://doi.org/10.3115/v1/D14-1162

Qu, J, Arguello, J., & Wang, Y. (2021). A Deep Analysis of an Explainable Retrieval Model for Precision Medicine Literature Search. *Proceedings of the 43rd European Conference on Information Retrieval (ECIR'21)*. ECIR, Lucca.

Qu, Jiaming, Arguello, J., & Wang, Y. (2020). Towards explainable retrieval models for precision medicine literature search. *Proceedings of the 43rd International ACM SIGIR Conference on Research and Development in Information Retrieval*, 1593–1596. https://doi.org/10.1145/3397271.3401277

Rivas, A. R., Iglesias, E. L., & Borrajo, L. (2014). Study of query expansion techniques and their application in the biomedical information retrieval. *Thescientificworldjournal*, *2014*, 132158. https://doi.org/10.1155/2014/132158

Rose, S., Engel, D., Cramer, N., & Cowley, W. (2010). Automatic Keyword Extraction from Individual Documents. In M. W. Berry & J. Kogan (Eds.), *Text mining: applications and theory* (pp. 1–20). Wiley. https://doi.org/10.1002/9780470689646.ch1

Roy, D., Paul, D., Mitra, M., Garain, U., & Roy, D. (2016, June 24). Using Word Embeddings for Automatic Query Expansion. *Neu-IR '16*. SIGIR Workshop on Neural Information Retrieval, Pisa, Italy.

Russell-Rose, T., Chamberlain, J., & Azzopardi, L. (2018). Information retrieval in the workplace: A comparison of professional search practices. *Information Processing & Management*, *54*(6), 1042–1057. https://doi.org/10.1016/j.ipm.2018.07.003

Russell-Rose, T., & Chamberlain, J. (2016a). Real-World Expertise Retrieval: The Information Seeking Behaviour of Recruitment Professionals. In *Advances in information retrieval*

(Vol. 9626, pp. 669–674). Springer. https://doi.org/10.1007/978-3-319-30671-1_51

Russell-Rose, T., & Chamberlain, J. (2016b). Searching for talent: The information retrieval challenges of recruitment professionals. *Business Information Review*, *33*(1), 40–48.

Russell-Rose, T., & Chamberlain, J. (2017). Expert search strategies: the information retrieval practices of healthcare information professionals. *JMIR Medical Informatics*, *5*(4), e33. https://doi.org/10.2196/medinform.7680

Ruthven, I. (2003). Re-examining the potential effectiveness of interactive query expansion. *SIGIR*, 213–220.

Scells, H., Zuccon, G., Koopman, B., & Clark, J. (2020, April 14). A Computational Approach for Objectively Derived Systematic Review Search Strategies. *Proceedings of ECIR 2020*. 42nd European Conference on Information Retrieval.

Scells, H., & Zuccon, G. (2018). searchrefiner: A Query Visualisation and Understanding Tool for Systematic Reviews. *Proceedings of the 27th ACM International Conference on Information and Knowledge Management  - CIKM '18*, 1939–1942. https://doi.org/10.1145/3269206.3269215

Seitner, J., Bizer, C., Eckert, K., Faralli, S., Meusel, R., Paulheim, H., & Ponzetto, S. P. (2016). A Large DataBase of Hypernymy Relations Extracted from the Web. *LREC*. LREC.

Shojania, K. G., Sampson, M., Ansari, M. T., Ji, J., Doucette, S., & Moher, D. (2007). How quickly do systematic reviews go out of date? A survival analysis. *Annals of Internal Medicine*, *147*(4), 224–233. https://doi.org/10.7326/0003-4819-147-4-200708210-00179

Song, W., Liang, J. Z., Cao, X. L., & Park, S. C. (2014). An effective query recommendation approach using semantic strategies for intelligent information retrieval. *Expert Systems with Applications*, *41*(2), 366–372. https://doi.org/10.1016/j.eswa.2013.07.052

Stefanowski, J., & Weiss, D. (2003). Carrot2 and language properties in web search results clustering. In *Advances in web intelligence* (pp. 240–249). Springer. https://doi.org/10.1007/3-540-44831-4_25

Tahery, S., & Farzi, S. (2020). Customized query auto-completion and suggestion—A review. *Information Systems*, *87*, 101415.

Tait, J. I. (2014). An introduction to professional search. In *Professional search in the modern world* (pp. 1–5). Springer.

Verberne, S., He, J., Kruschwitz, U., Larsen, B., Russell-Rose, T., & de Vries, A. P. (2018). First international workshop on professional search (profs2018). *SIGIR*, 1431–1434. https://doi.org/10.1145/3209978.3210198

Verberne, S., He, J., Kruschwitz, U., Wiggers, G., Larsen, B., Russell-Rose, T., & de Vries, A. P. (2019). First international workshop on professional search. *ACM SIGIR Forum*, *52*(1), 153–162. https://doi.org/10.1145/3308774.3308799

Verberne, S., Sappelli, M., & Kraaij, W. (2014). Query term suggestion in academic search. In *Advances in information retrieval* (Vol. 8416, pp. 560–566). Springer. https://doi.org/10.1007/978-3-319-06028-6_57

Verberne, S., Wabeke, T., & Kaptein, R. (2016, March 20). Boolean queries for news monitoring: Suggesting new query terms to expert users. *Proceedings of the NewsIR'16 Workshop at ECIR*. ECIR, Padua, Italy.

White, R. W., & Roth, R. A. (2009). Exploratory Search: Beyond the Query-Response Paradigm. *Synthesis Lectures on Information Concepts, Retrieval, and Services*, *1*(1), 1–98. https://doi.org/10.2200/S00174ED1V01Y200901ICR003

White, R. W. (2016). *Interactions with search systems* (1st ed.). Cambridge University Press.

Wu, F., Qiao, Y., Chen, J.-H., Wu, C., Qi, T., Lian, J., Liu, D., Xie, X., Gao, J., Wu, W., & Zhou, M.

(2020). MIND: A Large-scale Dataset for News Recommendation. *Proceedings of the 58th Annual Meeting of the Association for Computational Linguistics*, 3597–3606. https://doi.org/10.18653/v1/2020.acl-main.331

Wu, Y., Wu, W., Xing, C., Xu, C., Li, Z., & Zhou, M. (2019). A Sequential Matching Framework for Multi-Turn Response Selection in Retrieval-Based Chatbots. *Computational Linguistics*, *45*(1), 163–197. https://doi.org/10.1162/coli_a_00345

Xiong, C., & Callan, J. (2015). Query Expansion with Freebase. *ICTIR*, 111–120. https://doi.org/10.1145/2808194.2809446

Zeng, Q. T., Redd, D., Rindflesch, T., & Nebeker, J. (2012). Synonym, topic model and predicate-based query expansion for retrieving clinical documents. *AMIA Annual Symposium Proceedings*, *2012*, 1050–1059.